**Theoretical analysis of the inverse Edelstein effect at the LaAlO$_3$ / SrTiO$_3$ interface with an effective tight-binding model: Important role of the second $d_{xy}$ subband**

Shoma Arai[1], Shingo Kaneta-Takada[1], Le Duc Anh[1,2], Masaaki Tanaka[1,3], and Shinobu Ohya[1,2,3]*

[1]*Department of Electrical Engineering and Information Systems, The University of Tokyo, Hongo, Bunkyo-ku, Tokyo 113-8656, Japan*

[2]*Institute of Engineering Innovation, Graduate School of Engineering, The University of Tokyo, Bunkyo-ku, Tokyo 113-8656, Japan*

[3]*Center for Spintronics Research Network, Graduate School of Engineering, The University of Tokyo, Hongo, Tokyo 113-8656, Japan*

*E-mail: ohya@cryst.t.u-tokyo.ac.jp

The two-dimensional electron gas formed at interfaces between $SrTiO_3$ and other materials has attracted much attention since extremely efficient spin-to-charge current conversion has been recently observed at these interfaces. This has been attributed to their complicated quantized multi-orbital structures with a topological feature. However, there are few reports quantitatively comparing the conversion efficiency values between experiments and theoretical calculations at these interfaces. In this study, we theoretically explain the experimental temperature dependence of the spin-to-charge current conversion efficiency using an 8×8 effective tight-binding model considering the second $d_{xy}$ subband, revealing the vital role of the quantization of the multi-band structure.

In spintronics, developing material systems enabling highly efficient spin-charge conversion is strongly demanded for exploring low-power-consumption non-volatile devices, such as spin-orbit-torque magnetoresistive random access memory.[1] The two-dimensional electron gas (2DEG) formed at the SrTiO$_3$ (STO) interface, such as LaAlO$_3$ (LAO)/STO, is very promising because of its large Rashba spin-orbit interaction (SOI).[2,3] SOI causes spin splitting of the Fermi surface into the ones with different spin chirality, causing the spin-charge conversion in the 2DEG region [Fig. 1(a)]. The spin-to-charge conversion at material interfaces is called the inverse Edelstein effect (IEE),[4–8] and its conversion efficiency $j_c^{2D}/j_s$ is called the inverse Edelstein length $\lambda_{IEE}$, where $j_s$ is the injected spin current density and $j_c^{2D}$ is the generated two-dimensional current density in the 2DEG region. At the STO interface, previous studies have demonstrated significantly large $\lambda_{IEE}$ values up to ~60 nm, which are thought to originate from its characteristic band structure[9–16]: As shown in recent angle-resolved photoemission spectroscopy (ARPES) measurements, $d$-electrons are quantized due to band bending induced by the electric field near the interface in STO,[17] generating a complicated multi subband structure with a topological feature.[18] In particular, the second $d_{xy}$ subband seems to contribute to the IEE substantially.[19] However, there are few reports quantitatively comparing the $\lambda_{IEE}$ values between experiments and theoretical calculations based on the observed subband structure. Vaz *et al.* reported the experimental gate-voltage dependence of the $\lambda_{IEE}$ using the 2DEG formed at AlO$_x$/STO and compared it with a calculation result.[16] They calculated the Edelstein-effect efficiency using a semiclassical Boltzmann transport theory and a tight-binding model, which characterize the spin density as a response to an externally applied electric field.[16,18,20] Meanwhile, the effective tight-binding model proposed by Kim *et al.*[21] is convenient for calculating the $\lambda_{IEE}$, because the way to solve the eigen equations derived from this model has been well studied and established.[22)] In this study, we calculate the $\lambda_{IEE}$ considering the second $d_{xy}$ subband using the effective tight-binding calculation[16,19,23)] and compare the calculated result with previous experimental data obtained for the 2DEG at the LAO/STO interface [Fig. 1(b)].[24)] The experimental temperature dependence of the $\lambda_{IEE}$ is well reproduced by our theoretical calculation considering the second $d_{xy}$ subband with reasonable physical parameters. We find that the second $d_{xy}$ subband has a substantial contribution to the $\lambda_{IEE}$ in this material system.

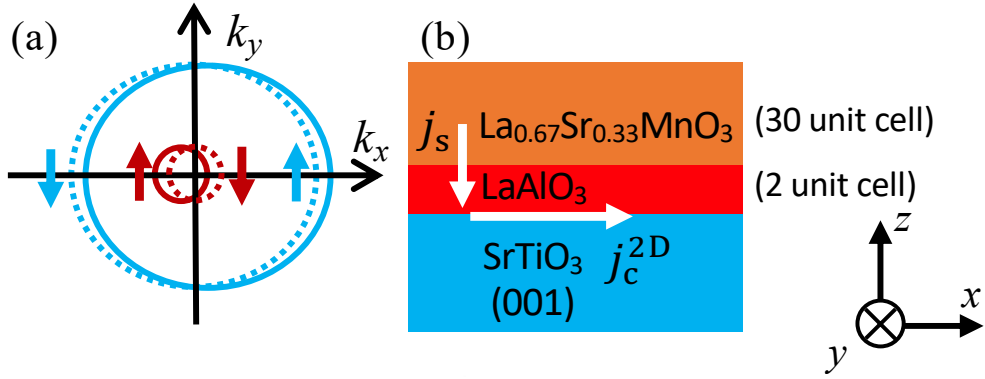

Fig. 1. (Color online) (a) Mechanism of the spin-to-charge current conversion via the inverse Edelstein effect in a simple parabolic-band picture. The spin current injected into the 2DEG region at the LAO/STO interface moves the outer and inner Fermi circles in opposite directions, generating a charge current in the $x$ direction. Here, the dotted and solid lines are the Fermi circles before and after the spin current is injected, respectively. (b) All-epitaxial oxide heterostructure used for the spin-to-charge current conversion experiment in our previous study.[24)] The spin current is injected from ferromagnetic La$_{0.67}$Sr$_{0.33}$MnO$_3$ (LaSrMnO$_3$) and converted to the charge current in the 2DEG region at the LAO/STO interface.

For the band calculation, we use the 8×8 Hamiltonian, for which we extend the 6×6 model proposed by Kim *et al.* by adding a basis of the second $d_{xy}$ subband.[21)] As shown in previous ARPES measurements, the second $d_{xy}$ subband is observed just below the Fermi level $E_F$ due to its relatively weak quantization, while the second and higher subbands of $d_{yz}$ and $d_{zx}$ do not appear below $E_F$ due to strong quantization.[16)] In the following, $H_0$ expresses onsite interaction (diagonal in orbital space) and nearest-neighbor hopping; $H_{ASO}$ represents spin-orbit interaction of atoms; $H_a$ represents nearest-neighbor hopping induced primarily by polar lattice distortion due to the electric field originating from broken inversion symmetry. Using the eight $t_{2g}$ orbitals of STO, $d_{yz\uparrow}$, $d_{yz\downarrow}$, $d_{zx\uparrow}$, $d_{zx\downarrow}$, $d_{xy1\uparrow}$, $d_{xy1\downarrow}$, $d_{xy2\uparrow}$, and $d_{xy2\downarrow}$ as basis functions, $H_0$, $H_{ASO}$, and $H_a$ are written as

$$H_0 = \begin{pmatrix} \frac{\hbar^2 k_x^2}{2m_h} + \frac{\hbar^2 k_y^2}{2m_l} & 0 & 0 & 0 \\ 0 & \frac{\hbar^2 k_x^2}{2m_l} + \frac{\hbar^2 k_y^2}{2m_h} & 0 & 0 \\ 0 & 0 & \frac{\hbar^2 k_x^2}{2m_l} + \frac{\hbar^2 k_y^2}{2m_l} - \Delta_{E1} & 0 \\ 0 & 0 & 0 & \frac{\hbar^2 k_x^2}{2m_l} + \frac{\hbar^2 k_y^2}{2m_l} - \Delta_{E2} \end{pmatrix} \otimes \sigma_0, \quad (1)$$

$$H_{ASO} = \Delta_{ASO} \begin{pmatrix} 0 & i\sigma_z & -i\sigma_y & -i\sigma_y \\ -i\sigma_z & 0 & i\sigma_x & i\sigma_x \\ i\sigma_y & -i\sigma_x & 0 & 0 \\ i\sigma_y & -i\sigma_y & 0 & 0 \end{pmatrix} \quad (2)$$

$$H_a = \Delta_z \begin{pmatrix} 0 & 0 & ik_x & ik_x \\ 0 & 0 & ik_y & ik_y \\ -ik_x & -ik_y & 0 & 0 \\ -ik_x & -ik_y & 0 & 0 \end{pmatrix} \otimes \sigma_0, \quad (3)$$

where $\hbar$ is the Dirac constant, $k_x$ and $k_y$ are the wavenumbers in the $x$ and $y$ directions, respectively, $\Delta_{E1}$ ($\Delta_{E2}$) represents the energy difference at the $\Gamma$ point ($k_x = k_y = 0$) between the first (second) $d_{xy}$ subband and the $d_{yz}$ band, which is lifted from the first $d_{xy}$ subband bottom due to the confinement of the wave function in the $z$ direction. $\Delta_{ASO}$ and $\Delta_z$ represent the magnitudes of $H_{ASO}$ and $H_a$, respectively. $\sigma_0$ is the identity matrix in spin space, and $\sigma_x$, $\sigma_y$, and $\sigma_z$ are the spin matrices. $\otimes$ is the Kronecker product. $m_l$ (= 0.41 $m_0$) and $m_h$ (= 6.8 $m_0$) are the effective masses of the light and heavy electrons at the LAO/STO interface, respectively, and $m_0$ is the free electron mass.[25] To check the influence of the second $d_{xy}$ subband on the $\lambda_{IEE}$, we also carry out the calculation using the 6×6 model without considering the basis of the second $d_{xy}$ subband. Figure 2 shows the band structure obtained from the Hamiltonian $H = H_0 + H_{ASO} + H_a$ when both $\Delta_{ASO}$ and $\Delta_z$ are set at 10 meV. Here, $a$ represents the lattice constant of STO. One can see that the red curve originating from the second $d_{xy}$ subband is added in Fig. 2(b) in comparison with the result obtained with the 6×6 model [Fig. 2(a)]. As shown in the enlarged view of the band structure in the inset of Fig. 2(b), we can see that each band is spin-split due to SOI.

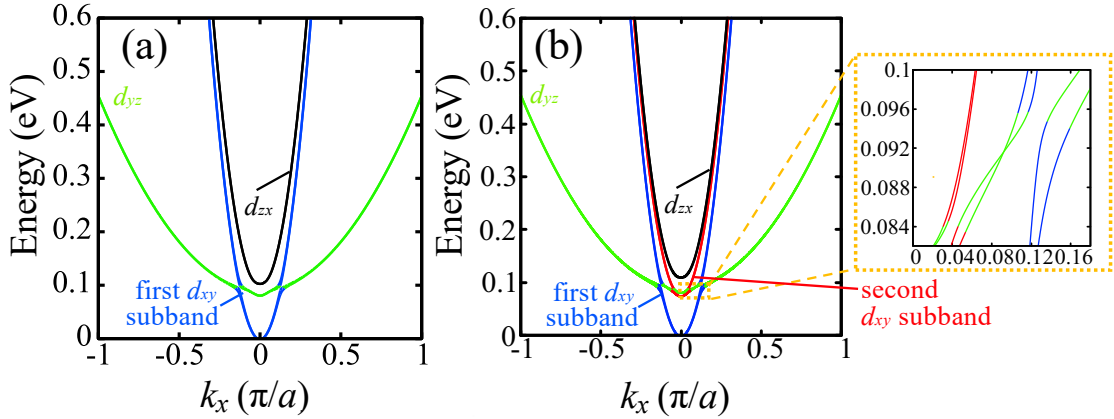

Fig. 2. (Color online) (a) Band structure obtained without considering the second $d_{xy}$ subband (6×6 model). (b) Band structure obtained with the second $d_{xy}$ subband (8×8 model). The black curve represents the $d_{zx}$ band, the red curve represents the second $d_{xy}$ subband, the blue curve is the first $d_{xy}$ subband, and the green curve is the $d_{yz}$ band. Here, $\Delta_z$ =10 meV, $\Delta_{ASO}$ = 10 meV, $\Delta_{E1}$ = 0.09 eV, and $\Delta_{E2}$ = 0.01 eV. The enlarged image shows the band structure near the band crossing point.

We calculate the $\lambda_{\text{IEE}}$ using the energy dispersion and the eigenstate derived from $H$ at each $\mathbf{k}$ [= $(k_x, k_y)$] point with fixed energy. Using the Boltzmann equation, $j_c^{2D}$ and $j_s$ are derived by the following equations:[26)]

$$j_c^{\text{FS}n} = \frac{e^2}{4\pi^2 \hbar} \int^{\text{FS}n} F_x(\mathbf{k}) dS_F, \tag{4}$$

$$\delta s^{\text{FS}n} = \frac{e}{4\pi^2 \hbar} \int^{\text{FS}n} |S_y(\mathbf{k})| dS_F, \tag{5}$$

$$j_c^{2D} = \sum_n j_c^{\text{FS}n}, \tag{6}$$

$$j_s = \sum_n \frac{e\, \delta s^{\text{FS}n}}{\tau}, \tag{7}$$

where $F_x(\mathbf{k})$ and $S_y(\mathbf{k})$ are defined as follows:

$$F_x(\mathbf{k}) = F\, \text{sgn}\left(S_y(\mathbf{k})\right) \tau(|\mathbf{k}|)\, v_x(\mathbf{k}) \frac{v_x(\mathbf{k})}{|v(\mathbf{k})|}, \tag{8}$$

$$S_y(\mathbf{k}) = F\, \tau(|\mathbf{k}|)\, \sigma_y(\mathbf{k}) \frac{v_x(\mathbf{k})}{|v(\mathbf{k})|}. \tag{9}$$

Here, $F$ is the electric field generated in the 2DEG region, $e$ is the electron charge, $\delta s^{\text{FS}n}$ is the spin accumulation, $\tau(|\mathbf{k}|)$ is the momentum relaxation time of the electron at each $\mathbf{k}$ point, $v(\mathbf{k}) = \left(v_x(\mathbf{k}), v_y(\mathbf{k})\right)$ is the group velocity, $\sigma_y(\mathbf{k})$ is the spin magnitude in the $y$ direction, and $dS_F$ is the infinitesimal area (= length in two dimensions) of the Fermi surface. In. Eqs. (4) and (5), we conduct the integration over the $n$th Fermi surface FS$n$. From Eqs. (4), (6), and (8), the relationship between $j_c^{2D}$ and $F$ is expressed as

$$j_c^{2D} = \sum_n \frac{e^2}{4\pi^2 \hbar} \int^{\text{FS}n} F\, \tau(|\mathbf{k}|)\, v_x(\mathbf{k}) \frac{v_x(\mathbf{k})}{|v(\mathbf{k})|}\, dS_F. \tag{10}$$

We define the averaged momentum relaxation time $\tau$ as

$$\tau = \left[\sum_n \int^{\text{FS}n} \tau(|\mathbf{k}|)\, dS_F\right] \Big/ \left(\sum_n \int^{\text{FS}n} dS_F\right). \tag{11}$$

Here, $\tau(|\mathbf{k}|)$ is assumed to be proportional to $|\mathbf{k}|$. Then, the value of $\tau$ is estimated from Eq. (10) with derived $v(\mathbf{k})$ and the experimental sheet resistance ($= F / j_c^{2D}$).

Figure 3 shows the calculated Fermi surface plotted with $v_x(\mathbf{k})$ and the spin direction $\langle \boldsymbol{\sigma} \rangle$ at $E_F = 140$ meV (measured from the first $d_{xy}$ subband bottom), where $\boldsymbol{\sigma}$ is defined as $(\sigma_x, \sigma_y, \sigma_z)$. Using the eigenvector $\mathbf{C_k}$ obtained at each $\mathbf{k}$ point, $\langle \boldsymbol{\sigma} \rangle$ is expressed as $\mathbf{C_k^\dagger} \boldsymbol{\sigma} \mathbf{C_k}$. Along the $k_x$ axis at $k_y = 0$, the Fermi surfaces from the outside correspond to the $d_{yz}$ band, first $d_{xy}$ subband, second $d_{xy}$ subband, and $d_{zx}$ band. As shown in Fig. 3(a), at around $k_y = 0$, the second $d_{xy}$ subband has the next largest $v_x$ after the first $d_{xy}$ subband. As explained later, this result suggests that the second $d_{xy}$ subband largely contributes to the IEE. Although each band splits into the outer and inner Fermi surfaces, this split is small and is not distinguishable on this scale. Thus, we plot the spin split outer and inner Fermi surfaces of each band separately in the left and right side of Fig. 3(b), respectively. In both figures, at $k_y = 0$, starting from the outer Fermi surface, the spin directions are $|\uparrow\rangle$, $|\downarrow\rangle$, $|\downarrow\rangle$, and $|\uparrow\rangle$. This order agrees with a previous result obtained by a density functional theory calculation.[27]

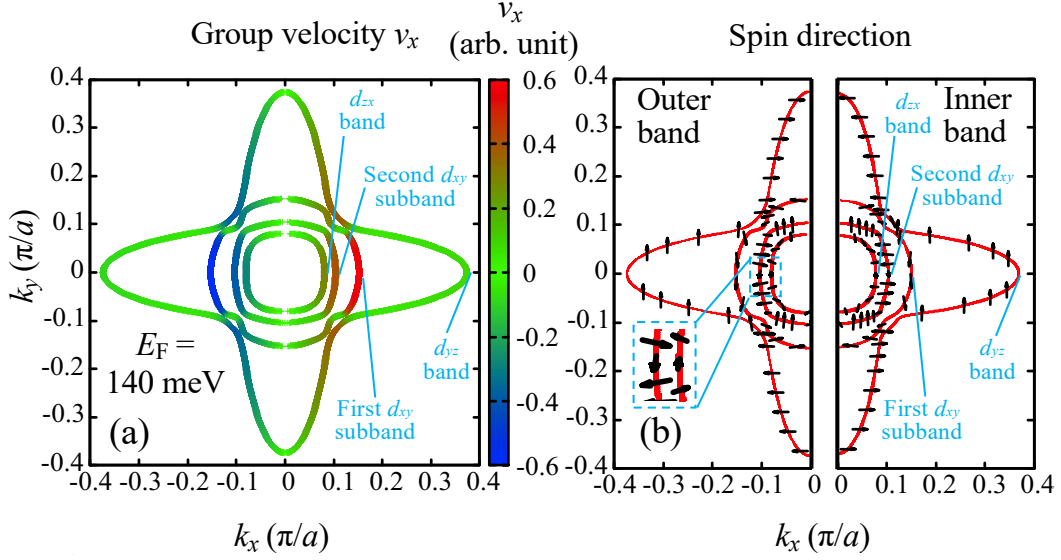

Fig. 3. (Color online) (a) Fermi surface mapped with the group velocity $v_x$ in the $x$ direction at $E_F$ = 140 meV. (b) Fermi surface mapped with the spin direction at each **k** point at $E_F$ = 140 meV. In (b), the left (right) side shows only the outer (inner) band. Here, the Fermi surface at $E_F$ = 140 meV is shown as an example because one can easily understand that band crossing generates the complicated Fermi surface near this energy.

Figure 4 shows the $E_F$ dependence of $j_c^{2D}/\delta s$ [= $(\tau/e) \times (j_c^{2D}/j_s)$, see Eqs. (6) and (7)] derived when the second $d_{xy}$ subband is taken into account (red) and when it is not taken into account (blue). Here, $\delta s$ is the summation of $\delta s^{FSn}$ over all $n$. One can see that $j_c^{2D}/\delta s$ increases by incorporating the second $d_{xy}$ subband when $E_F$ is above ~0.95 eV, at which the band bottom of the second $d_{xy}$ band is located. Because $j_c^{2D}$ is obtained by integrating $F_x(\mathbf{k})$ as shown in Eqs. (4) and (6), and because $F_x(\mathbf{k})$ is proportional to $v_x(\mathbf{k})$ as shown in Eq. (8), the charge current converted from the spin current is mainly composed of electrons with large $v_x(\mathbf{k})$. Thus, from Fig. 3(a), one can see that the first and second $d_{xy}$ subbands have the main contribution to the $\lambda_{IEE}$. This is the most probable reason for the increase in the $\lambda_{IEE}$ by considering the second $d_{xy}$ subband.

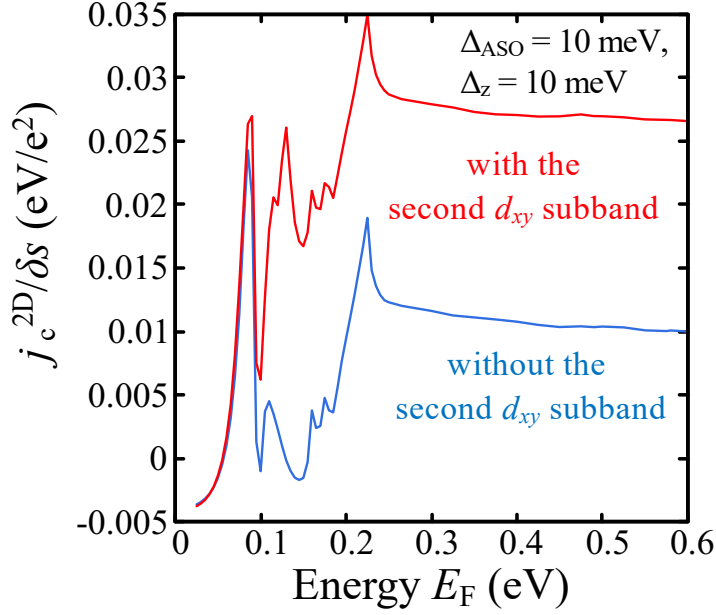

Fig. 4. (Color online) $E_F$ dependence of $j_c^{2D}/\delta s$ when the second $d_{xy}$ subband is taken into account (red line) when it is not (blue line).

In the following, we compare the experimental temperature dependence and calculation results of the $\lambda_{IEE}$ using the 8×8 model. Here, we approximate that the IEE occurs only at the $E_F$ because the signals were mainly obtained at low temperatures less than 150 K in the experiment.[24] We determine the $E_F$ value at each temperature so that the carrier density $n$ calculated by the Green function method using the band dispersion shown in Fig. 2(b)[28] equals the experimental one; $(E_F, n) = $ (210 meV, 2.13 × 10$^{14}$ cm$^{-2}$), (220 meV, 2.21 × 10$^{14}$ cm$^{-2}$), (265 meV, 2.90 × 10$^{14}$ cm$^{-2}$), (375 meV, 4.64 × 10$^{14}$ cm$^{-2}$), (490 meV, 6.39 × 10$^{14}$ cm$^{-2}$), and (580 meV, 7.43 × 10$^{14}$ cm$^{-2}$) at 20, 40, 60, 80, 100, and 140 K, respectively.

Figure 5 shows the comparison between the theoretical and experimental results of the $\lambda_{IEE}$ as a function of temperature. The calculated results show good agreement with the experimental results for $\Delta_{ASO} = 20$ meV when the second $d_{xy}$ subband is not considered and for $\Delta_{\text{ASO}}$ = 10 meV when the second $d_{xy}$ subband is considered. Here, we discuss the value of the effective Rashba coefficient $\alpha_{\text{eff}} = \hbar^2 \Delta k/(2m)$, where we define $\Delta k$ as maximum band spin-splitting in the $k_x$ direction at $k_y = 0$ for the first $d_{xy}$ subband (*i.e.* $E_F$ = 85 meV). When $\Delta_z$ = 10 meV and $\Delta_{\text{ASO}}$ = 10 meV, $\alpha_{\text{eff}}$ is estimated to be $8.15 \times 10^{-13}$ eVm when we incorporate the second $d_{xy}$ subband, which almost agrees with the reported values of $\alpha_{\text{eff}}$ (~$1.2 \times 10^{-12}$ eVm) for LAO/STO.[3] For comparison, we calculate $\alpha_{\text{eff}}$ using the 6×6 model. $\alpha_{\text{eff}}$ is estimated to be $1.48 \times 10^{-12}$ eVm, which is larger than that obtained with the 8×8 model. Both 6×6 and 8×8 models can reproduce the temperature dependence of the $\lambda_{\text{IEE}}$; however, since the second $d_{xy}$ subband is observed by ARPES measurements,[16] we need to incorporate the second $d_{xy}$ subband.

Our results show that the experimental results of the temperature dependence of the $\lambda_{\text{IEE}}$ can be well explained by our theoretical calculation considering the second $d_{xy}$ subband with reasonable physical parameters. As shown in Fig. 4, this calculation indicates that the quantization of the $d_{xy}$ band has a significant influence on the spin-to-charge conversion, suggesting that we can enhance the conversion efficiency by controlling the quantization of interfacial bands with the same spin chirality. In particular, controlling the quantum confinement for each band will be vital for such multiband materials like STO. Moreover, in Fig. 3 (b), one can see characteristic topologically non-trivial avoided anti-crossing points between the second $d_{xy}$ subband and $d_{yz}$ band at around $(k_x, k_y) = (\pm 0.085\ \pi/a, \pm 0.085\ \pi/a)$, where the spin expectation value almost vanishes (the length of the spin arrows becomes nearly zero). This characteristic feature can be attributed to the large spin-orbit interaction of this system.[16] Because the electron states near these points have finite $v_x(\mathbf{k})$, they largely contribute to the $\lambda_{\text{IEE}}$, which suggests the importance of quantization in systems with large SOI. Our findings will help develop and design material interfaces with large spin-charge conversion in the future.

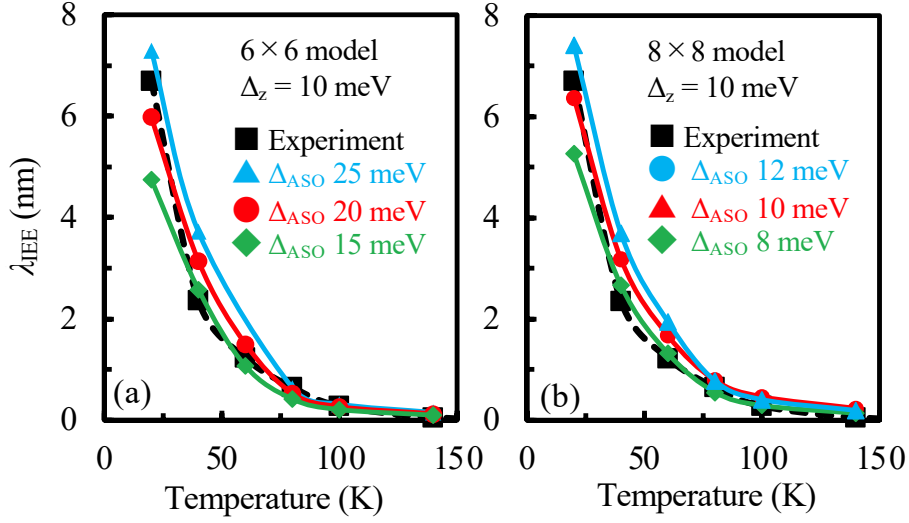

Fig. 5. (Color online) (a) and (b) Temperature dependence of the experimental (black)[24] and calculated $\lambda_{IEE}$ values (blue, red, and green curves). The calculated $\lambda_{IEE}$ values in (a) represent the results obtained without considering the second $d_{xy}$ subband (6×6 model). The ones in (b) represent the results obtained when considering the second $d_{xy}$ subband (8×8 model).

In summary, to explain the temperature dependence of the $\lambda_{IEE}$ obtained at the LAO/STO interface, we have performed a theoretical calculation incorporating the second $d_{xy}$ subband using an effective tight-binding model. The calculation result considering the second $d_{xy}$ subband reproduces the experimental results well with the reasonable parameters. We have found that the second $d_{xy}$ subband has a large contribution to the $\lambda_{IEE}$. Our study gives us a clue to search for materials that exhibit highly efficient spin-charge conversion and will lead to a deeper understanding of this phenomenon.

**ACKNOWLEDGEMENTS**

This work was partly supported by Grants-in Aid Scientific Research (18H03860), the CREST program of Japan Science and Technology Agency (JPMJCR1777), and Spintronics Research Network of Japan.

**REFERENCES**

1. F. Oboril, R. Bishnoi, M. Ebrahimi, and M. B. Tahoori, IEEE Trans. Comput.-Aided Design Integr. Circuits Syst. **34**, 367 (2015).

2. A. Ohtomo and H. Y. Hwang, Nature **427**, 423 (2004).

3. A. D. Caviglia, M. Gabay, S. Gariglio, N. Reyren, C. Cancellieri, and J.-M. Triscone, Phys. Rev. Lett. **104**, 126803 (2010).

4. J. C. Rojas-Sánchez, L. Vila, G. Desfonds S. Gambarelli, J. P. Attané, J. M. De Teresa, and A. Fert, Nat. Commun. **4**, 2944 (2013).

5. P. Noel, C. Thomas, Y. Fu, L. Vila, B. Haas, P-H. Jouneau, S. Gambarelli, T. Meunier, P. Ballet, and J. P. Attanè, Phys. Rev. Lett. **120**, 167201 (2018).

6. S. Oyarzún, A. K. Nandy, F. Rortais, J. C. Rojas-Sánchez, M. T. Dau, P. Noël, and M. Jamet, Nat. Commun. **7**, 13857 (2016).

7. J.-C. Rojas-Sánchez, S. Oyarzún, Y. Fu, A. Marty, C. Vergnaud, S. Gambarelli, L.Vila, M. Jamet, Y. Ohtsubo, A. Taleb-Ibrahimi, P. Le Fèvre, F. Bertran, N. Reyren, J.-M. George, and A. Fert, Phys. Rev. Lett. **116**, 096602 (2016).

8. J.-C. Rojas-Sànchez and A. Fert, Phys. Rev. Appl. **11**, 054049 (2019).

9. E. Lesne, Y. Fu, S. Oyarzun, J. C. Rojas-Sánchez, D. C. Vaz, H. Naganuma, G. Sicoli, J. P. Attané, M. Jamet, E. Jacquet, J. M.George, A. Barthélémy, H. Jaffrès,

A. Fert, M. Bibes, and L.Vila, Nat. Mater. **15**, 1261 (2016).

10. Q. Song, H. Zhang, T. Su, W. Yuan, Y. Chen, W. Xing, J. Shi, J. Sun, and W. Han, Sci. Adv. **3**, e1602312 (2017).

11. Y. Wang, R. Ramaswamy, M. Motapothula, K. Narayanapillai, D. Zhu, J. Yu, T. Venkatesan, and H. Yang, Nano Lett. **17**, 7659 (2017).

12. J. Y. Chauleau, M. Boselli, S. Gariglio, R.Weil, G. De Loubens, J. M. Triscone, and M. Viret, Europhys. Lett. **116**, 17006 (2016).

13. W. Zhang, Q. Wang, B. Peng, H. Zeng, W. T. Soh, C. K. Ong, and W. Zhang, Appl. Phys. Lett. **109** 262402 (2016).

14. G. Seibold, S. Caprara, M. Grilli, and R. Raimondi, Phys. Rev. Lett. **119**, 256801 (2017).

15. P. Noël, F. Trier, L. M. V. Arche, J. Bréhin, D. C. Vaz, V. Garcia, S. Fusil, A. Barthélémy, L. Vila, M. Bibes and J. P. Attané, Nature **580**, 483 (2020).

16. D. C. Vaz, P. Noël, A. Johansson, B. Göbel, F. Y. Bruno, G. Singh, S. M. K. Walker, F. Trier, L. M. V. Arche, A. Sander, S. Valencia, P. Bruneel, M. Vivek, M. Gabay, N. Bergeal, F. Baumberger, H. Okuno, A. Barthélémy, A. Fert, L. Vila, I. Mertig, J. P. Attané, and M. Bibes, Nat. Mater. **18**, 1187 (2019).

17. A. F. Santander-Syro, O. Copie, T. Kondo, F. Fortuna, S. Pailhés, R. Weht, X. G. Qiu, F. Bertran, A. Nicolaou, A. Taleb-Ibrahimi, P. Le Fèvre, G. Herranz, M. Bibes, N. Reyren, Y. Apertet, P. Lecoeur, A. Barthélémy, and M. J.Rozenberg, Nature **469,** 189 (2011).

18. M. Vivek, M. O. Goerbig, and M. Gabay, Phys. Rev. B **95**, 165117 (2017).

19. P. Bruneel and M. Gabay, Phys. Rev. B **102**, 144407 (2020).

20. K. Shen, G. Vignale, and R. Raimondi, Phys. Rev. Lett. **112**, 096601 (2014).

21. Y. Kim, R. M. Lutchyn, and C. Nayak, Phys. Rev. B **87**, 245121 (2013).
22. Y. X. Liu, D. Z. -Y. Ting, and T. C. McGill, Phys. Rev. B **54**, 5675 (1996).
23. A. Johansson, B. Göbel, J. Henk, M. Bibes, and I. Mertig, Phys. Rev. Res. **3**, 013275 (2021).
24. S. Ohya, D. Araki, L. D. Anh, S. Kaneta, M. Seki, H. Tabata, and M. Tanaka, Phys. Rev. Res. **2**, 012014(R) (2020).
25. Z. Zhong, A. Tóth, and K. Held, Phys. Rev. B **87**, 161102(R) (2013).
26. P. Gambardella and I. M. Miron, Philos. Trans, R. Soc., A **369**, 3175 (2011).
27. P. D. C. King, S. M. K. Walker, A. Tamai, A. De La Torre, T. Eknapakul, P. Buaphet, S. K. Mo, W. Meevasana, M. S. Bahramy, and F. Baumberger, Nat. Commun. **5**, 3414 (2014).
28. H. Bahlouli, A. D. Alhaidari, and M. S. Abdelmonem, Phys. Lett. A, **367**, 162 (2007).